\documentstyle[11pt,aaspp4,flushrt]{article}

\begin{document}

\title{To be appeared in ApJ 490 (1997) 
\\
An equation of state from cool-dense fluids to hot gases for mixed elements}

\author{G.  Q.  Luo \altaffilmark{1}}

\altaffiltext{1}{Postdoctor, Beijing astronomical observatory.}

\affil{Yunnan Observatory, Kunming, 650011, P.R.  China\\
Beijing astronomical observatory, Beijing, 100080, P.R. China}

\authoremail{luogq@bao01.bao.ac.cn}

\baselineskip = 12pt

\begin{abstract}
An equation of state for the domain extending from hot gases to cool-dense
fluids is formulated for a hydrogen-helium mixture. The physical processes
take account of temperature ionization and dissociation, electron
degeneracy, Coulomb coupling and pressure ionization. Pressure ionization
and Coulomb coupling are studied with simple and comprehensive modeling. A
single and complete algorithm is achieved with explicit expressions
available for the whole domain from hot gases to cool dense fluids ($T>10^2%
{\rm K}$). Pressure ionization and Coulomb coupling have been examined for
their contributions to the pressure and internal energy. The result reveals
that their contributions smooth the variation of the pressure and internal
energy in the region of pressure ionization even at very low temperatures.
\end{abstract}


\section{Introduction}

The equation of state (EOS) of cool dense fluids is concerned with the
interiors of very low mass stars, brown dwarfs and giant planets in
astrophysics. The complexity of the physical processes of such an EOS stems
from the inclusion of interparticle interactions among atoms, molecules,
ions and electrons. The physical processes concerned mainly include two
types, i.e. pressure ionization and Coulomb coupling. Many studies have been
carried on for Coulomb coupling among charged particle. For instance, the
internal energy of fully ionized plasmas of classical one-component has been
given by advanced physics models (see Slattery and Doolen 1980) for the case
of strong coupling. The free energy of fully ionized hydrogen-helium mixture
has also been evaluated for thermodynamic properties with strong coupling
(see DeWitt and Hubbard 1976). The Coulomb coupling in weak limits can be
well described by Debye-Huckel model (see Cox and Giuli 1968; Mihalas et al.
1988). However, the Coulomb coupling between strong and weak limits still
remains for the study of its model and effects. In addition, the problem of
pressure ionization has been remaining thorny in the EOS calculation for
cool dense fluids (see Saumon et al. 1995). Although the advanced physical
models of pressure ionization have been studied for the EOS (see Blenski and
Ishikawa 1995; Saumon and Chabrier 1991, 1992), it is not easy to
incorporate them into the astrophysical EOS for their practical use and
investigation. In this paper, we establish the physical models for pressure
ionization and Coulomb coupling which have simple expressions and can be
easily incorporated into the EOS formalism.

Almost all of the EOS for cool dense fluids are given with their results in
tabular form for astrophysical application (e.g. Fontain et al. 1977; Magni
and Mazzitelli 1979; Mihalas et al. 1988; Saumon et al. 1995). This is
because the complex formalism and time-consuming calculation of the EOS for
cool dense fluids stem from many physical processes. They include
temperature dissociation and ionization, electron degeneracy, Coulomb
coupling and pressure ionization, etc. However, explicit and simplified
expressions for a comprehensive EOS can be of benefit to the study of the
unfamiliar physical processes concerned in the EOS and to the study of their
importance to and influences on astrophysical phenomena. We aim to present
such an algorithm for the EOS covering the domain from cool dense fluids to
hot gases.

Furthermore, it is also important to assure the reliability for the
formulated EOS and its results. Therefore, we not only establish the
explicit and simple expressions for the EOS but also compare the resulted
EOS with those from other more sophisticated models. In order to preserve
the continuity and generality of the EOS, we formulate the EOS on the basis
of the grand canonical approach in the chemical picture for the EOS (see Luo
1994). It allows to establish a single algorithm of the EOS available for
the whole domain from cool dense fluids to hot gases, even when many
physical processes are taken into account. The grand canonical approach in
the chemical picture also allows to include mixed elements in the EOS. The
EOS presented in this paper is formulated for a hydrogen(H)-helium(He)
mixture and takes into account the physical processes of atomic ionization,
molecular dissociation, electron degeneracy, Coulomb coupling, pressure
ionization, etc.

In Sect.2 we establish the models of pressure ionization and Coulomb
coupling. Sect.3 presents an algorithm of calculating thermal equilibrium.
The algorithm of calculating the pressure and internal energy is presented
in Sect. 4. In Sect. 5, the calculated results of the EOS are compared with
other EOS, and are examined for the contributions of pressure ionization and
Coulomb coupling to thermodynamic functions.

\section{Pressure ionization and Coulomb coupling}

We start from a H-He mixture gas with mass density $\rho $, temperature $T$
and the mass abundances $X$ and $Y$ for H and He. In a volume $V$, the
number of nuclei of H and He are given by 
\begin{equation}
N_{{\rm H}}=N_{{\rm A}}\rho VX/A_{{\rm H}},~~~~~N_{{\rm He}}=N_{{\rm A}}\rho
VY/A_{{\rm He}}, 
\end{equation}
with the Avagadro's number $N_{{\rm A}}$, the atomic weights $A_{{\rm H}%
}=1.0079$ and $A_{{\rm He}}=4.0026$.

Two physical processes, i.e., Coulomb coupling of charged particles and
pressure ionization of atoms, have been taken into account for the nonideal
effects on the EOS. Owing to the complexity of such physical processes,
their incorporation into the EOS requires some assumptions to allow
simplified and approximate models.

\subsection{Coulomb coupling}

The Coulomb coupling has considerable influences on the EOS, especially in
fully ionized plasmas. In partially ionized fluids, it plays a subtle role
in the determination of the ionization degrees as well as thermodynamic
quantities. We first approximate the Coulomb coupling by a one-component
model of fully ionized plasmas, and then add the corrections of partial
ionization and mixed elements to its formalism.

The Coulomb coupling can well be described in two limit cases, i.e., weak
coupling ($\Gamma \ll 1$) and strong coupling ($\Gamma \gg 1$). The Coulomb
coupling parameter $\Gamma =(Ze)^2/\bar rkT$ is defined by the nuclear
charge $Z$, the ion sphere radius $\bar r$ and temperature $T$. For a H-He
mixture, we define the average nuclear charge $Z$ as 
\begin{equation}
Z=\frac{N_{{\rm H}}+2N_{{\rm He}}}{N_{{\rm H}}+N_{{\rm He}}}. 
\end{equation}
The ion sphere radius $\bar r$ is determined by the number of ions $N_{{\rm %
ion}}$ in the volume $V$ using the approximation $(4\pi /3)\bar r^3N_{{\rm %
ion}}/V=1$. Considering the correction from the partial ionization, we
approximate the ion number by $N_{{\rm ion}}=N_{{\rm e}}/Z$ with the number
of free electrons $N_{{\rm e}}$. As a result, the Coulomb coupling parameter
can be written as 
\begin{equation}
\Gamma ={\frac{(Ze)^2}{kT}}\left( \frac{4\pi N_{{\rm e}}}{3ZV}\right) ^{1/3} 
\end{equation}
with the corrections from the partial ionization and mixed elements.

In the case of strong coupling ($\Gamma \gg 1$), the energy of a fully
ionized plasma with one component (of nuclear charge $Z$) is found to be 
\begin{equation}
E_{{\rm cc}}=-kTN_{{\rm ion}}\times 0.89752\Gamma ,\ \qquad (\Gamma \gg 1) 
\end{equation}
with the ion number $N_{{\rm ion}}$ (see Slattery et al. 1980). For weak
coupling $\Gamma \ll 1$, the Debye-H\"uckel model can be used to obtain its
Coulomb coupling energy (see Cox \& Giuli 1968). We ignore the effect of
electron degeneracy and suppose a fully ionized plasma with one component
(of nuclear charge $Z$). We thus have 
\begin{equation}
E_{{\rm cc}}=-kTN_{{\rm ion}}\times \left( 3/4\right) ^{1/2}\left[ \left(
1+1/Z\right) \Gamma \right] ^{3/2},\qquad 
 (\Gamma \ll 1). 
\end{equation}
The results for both the limit cases indicate that the Coulomb coupling
energy per ion in unit $kT$ (i.e. $E_{{\rm cc}}/N_{{\rm ion}}kT)$ is a
function of the Coulomb coupling parameter $\Gamma $ and the nuclei charge $%
Z $. The functions ($E_{{\rm cc}}/N_{{\rm ion}}kT$) of the two limit cases
for $Z=1$ are plotted versus $\Gamma $ in Fig. 1 which shows their
intersection at $\Gamma =0.13$.

In our EOS formalism the average potential energy of each free electron from
Coulomb coupling $\epsilon _{{\rm cc}}$ is adopted. The internal energy from
the Coulomb coupling is calculated in terms of 
\begin{equation}
E_{{\rm cc}}=N_{{\rm e}}\epsilon _{{\rm cc}}-N_{{\rm e}}\left( \frac{%
\partial \epsilon _{{\rm cc}}}{\partial \ln T}\right) _{VN} 
\end{equation}
which is the result of the EOS with the grand canonical approach in the
chemical picture (see Luo 1994, 1996). With this definition and the
constraints of the results of Coulomb coupling energy in two limit cases,
the average energy of each free electron from Coulomb coupling can be fitted
to be 
\begin{equation}
\epsilon _{{\rm cc}}=-kT\frac{g(\Gamma ,Z)}Z 
\end{equation}
with a fitting function $g(\Gamma ,Z)$ defined by 
\begin{eqnarray}
1/ g & = & 1 / g_1 + 1 / g_2 \;,
\\
g_1 & = &  (2/3) (3/4)^{1/2} (1+1/Z)^{3/2} \Gamma^{3/2} \;,
\nonumber\\
g_2 & =  & 0.89752 \Gamma \;.
\nonumber
\end{eqnarray}

The internal energy from Coulomb coupling resulted from the fitting function
is therefore given by 
\begin{equation}
E_{{\rm cc}}(\Gamma ,Z)=-kTN_{{\rm ion}}\times g(\Gamma ,Z)\left( \frac{%
\partial \ln g}{\partial \ln \Gamma }\right) . 
\end{equation}
according to Eqs. (3), (6) and (7). It has asymptotic form as in equation
(4) for strong coupling case and as equation (5) for weak coupling case. The
results from equation (9) for the Coulomb coupling energy per ion in unit $%
kT $ ($E_{{\rm cc}}/N_{{\rm ion}}kT$) with $Z=1$ are plotted in Fig. 1. It
is compared with the results of the two limit cases. The accurate results
for $\Gamma \ge 0.3$ (see Slattery and Doolen 1980) are also plotted in Fig.
1. They indicate that the simply fitted formula still overestimate the
Coulomb coupling energy about tens percent in the intermediate range of $%
\Gamma $. However, it is easy to add the power weighting to the fitting
formulae to obtain better fitted results.

\begin{figure}[h]
\vbox to 2in{\rule{0pt}{0in}}
\includegraphics{fig_1.eps}
\caption{\small
The fitting function for logarithmic ($E/kTN_{\rm{ion}}$) versus
logarithmic $\Gamma $ (solid line) , 
with those of the weak Coloumb coulping $\Gamma \ll 1$ (dotted line) 
and of the strong Coloumb coupling $\Gamma \gg 1$ (dashed line), 
and with the accurate results for $\Gamma >0.3$ (circles) given by Slattery and Doolen
(1980).}
\end{figure}

The fitted formula of equation (7) has a derivative with respect to $\ln
\Gamma $ 
\begin{equation}
{\frac{\partial \ln g}{\partial \ln \Gamma }}={\frac 52}-{\frac{(3/2)g_1+g_2 
}{g_1+g_2}}\;, 
\end{equation}
with the functions $g_1$ and $g_2$ defined in equation (8). It has
asymptotic values $3/2$ for $\Gamma \ll 1$ and $1$ for $\Gamma \gg 1$. It
will be used in Sect.4 for calculating the pressure and internal energy from
Coulomb coupling.

\subsection{Pressure ionization}

Pressure ionization is caused by the interaction of an atom with surrounding
particles which is responsible for the dissolution of atomic energy states.
In the gases with high temperatures (e.g. $T\ge 10^5K$), a great fraction of
free electrons and ions lead to the Stark ionization of atoms. This has been
well described by the MHD EOS with the occupation probability formalism (see
Hummer \& Mihalas 1988). For cool dense fluids, however, neutral atoms and
molecules dominate the interaction on an acted-upon atom. Accordingly, the
sophisticated physical models are established to produce continuous
solutions of the EOS (e.g. Saumon \& Chabrier 1992; Blenski \& Ishikava
1995). These models are based on the advanced physics theory and partly
examined by experimental data. For astrophysical application and research,
however, we may make approximate and statistical models for the complicated
physical processes of pressure ionization. This will allow to examine the
uncertainty of pressure ionization and its influence on the models of
astrophysical objects through the EOS.

Our model for pressure ionization is based on the energy change of an acted
atom by the pressure impact of acting particles. When an acted-upon atom is
collided by other particles, the pressure impact can always increase its
energy. If the acted-upon atom is supposed to have an effective volume $v_1$%
, the impact by acting particles with an equivalent pressure $P_1$ will
increase the energy of the acted-upon atom by an amount of $\Delta
E_1=P_1v_1 $. This is the result from the view of statistical mechanics. In
thermodynamic equilibrium with temperature $T$, the energy change $\Delta
E_1 $ leads to a Boltzmann factor $W=\exp (-\Delta E_1/kT)$ for the
existence probability of the atom. We thus can use the resulted factor $W$
to present the occupation probability for the specified atom.

The acting particles include all classical particle with the number of
atomic neclei $N_{{\rm H}}+N_{{\rm He}}$. They also include the bound
electrons of other atoms beyond the mean separation $d$ that the acted-upon
atom can feel. The mean separation of atoms $d$ is defined by 
\begin{equation}
\frac{4\pi }3d^3\left( \frac{N_{{\rm H}}+N_{{\rm He}}}V\right) =1 
\end{equation}
for a H-He mixture with the number of atomic nuclei $N_{{\rm H}}+N_{{\rm He}%
} $.

Suppose the atoms of H and He have the hydrogen-like wave functions of
ground states $w(Zr)=\sqrt{2}Zre^{-2Zr}$, we then have the number of the
bound electrons beyond the mean separation $d$ 
\begin{equation}
N_{{\rm e}}^{{\rm {b}}}=N_{{\rm H}}\int_d^\infty w(r){\rm d}r+2N_{{\rm He}%
}\int_d^\infty w(2r){\rm d}r=N_{{\rm H}}f(d)+2N_{{\rm He}}f(2d), 
\end{equation}
with a function $f(x)$ defined by 
\begin{equation}
f(x)=e^{-2x}(1+2x+2x^2). 
\end{equation}

It is a rough approximation that the mean separation of atoms is used to
determine the number of the acting electrons. Because an atom with a kinetic
energy $(3/2)kT$ can always approach another atom closer than the mean
separation $d$, it can feel more bound electrons of other atoms than those
outside the separation $d$. As a result, the integration limit $d$ in
equation (12) should be corrected by taking account of the thermal motion of
the atom. Nevertheless, the thermal motion of the atom play a role in the
determination of the integration limit at relatively high temperatures. In
that case, temperature ionization play a more important role than pressure
ionization. We find that the inclusion of the thermal motion dose not have a
very important effect on the model of pressure ionization in the EOS
formalism. In addition, its inclusion will complicate the model of pressure
ionization in the EOS formalism. We therefore do not present its formulation
in this paper.

The acting atoms and ions are classical particles while the acting electrons
with the number $N_{{\rm e}}^{{\rm b}}$ behave like the Fermi-Dirac
particles. As a result, the equivalent pressure $P_1$ from the acting
particles is given by the Fermi-Dirac statistics 
\begin{equation}
P_1=kT{\left( \frac{{N_{{\rm H}}+N_{{\rm He}}}}V\right) +\frac 23}kT{\frac{%
N_{{\rm e}}^{{\rm b}}}V}{\frac{F_{3/2}(\lambda _1)}{F_{1/2}(\lambda _1)}} 
\end{equation}
with the Fermi-Dirac integral $F_{3/2}$ and $F_{1/2}$. The degeneracy of the
acting electrons $\lambda _1$ is determined by 
\begin{equation}
N_{{\rm e}}^{{\rm b}}/V=C_{{\rm e}}T^{3/2}F_{1/2}(\lambda _1) 
\end{equation}
with a constant $C_{{\rm e}}=(\sqrt{2}/\pi ^2)(km_{{\rm e}}/\hbar ^2)^{3/2}$
(see Luo 1994).

The effective volume $v_1$ of the acted-upon atom is defined by its radius $%
r_j$ in terms of $v_1=(4\pi /3)r_j^3$. As a result, we obtain the occupation
probability in terms of $W=\exp (-P_1v_1/kT)$ in a logarithmic form 
\begin{equation}
\ln W_j=-{\frac{4\pi }3}r_j^3\left[ {\frac{N_{{\rm H}}+N_{{\rm He}}}V}+{%
\frac 23}C_{{\rm e}}T^{3/2}F_{3/2}(\lambda _1)\right] 
\end{equation}
for an atom with an effective radius $r_j$. In equation (16) the first term
is identical to the result of the hard sphere model of atoms (see Hummer and
Mihalas 1988). The second term is the result of the interaction from the
bound electrons of other atoms on the acted-upon atom. The radii of atomic
hydrogen (HI) and singly ionized helium (HeII) are taken to be $r_{{\rm HI}%
}=a_0$ and $r_{{\rm HeII}}=0.5a_0$ with the Bohr radius $a_0$, and that of
atomic He is assumed to be $r_{{\rm HeI}}=0.8a_0$. The resulted occupation
probabilities are thereby represented by $W_{{\rm HI}}$, $W_{{\rm HeI}}$ and 
$W_{{\rm HeII}}$.

The occupation probability given by equation (16) is an explicit function of 
$T$, $N_{{\rm H}}$ and $N_{{\rm He}}$ as well as $r_j$. It thus can be
incorporated into the calculation of the thermal equilibrium. Its simplified
form allows to obtain analytical expressions for their contribution from
pressure ionization to the pressure and internal energy (see Sect. 4.5).

\section{Thermal equilibrium quantities of the EOS}

\subsection{Thermal equilibrium quantities}

Thermal equilibrium quantities are explicitly defined by the degrees of atom
ionization and molecule dissociation. For the EOS in the chemical picture,
the determination of thermodynamic functions significantly depends on the
degrees of ionization and dissociation. According to the grand canonical
approach in the chemical picture (see Luo 1994), the thermal equilibrium of
a mixture can be specified by the probabilities of atomic nuclei in various
configurations.

We total include six configurations for H and He. They are specified for
hydrogen by molecular state ${\rm H_2}$ (abbreviated as M), atomic state HI
and ionized state HII, as well as for helium by atomic state HeI, singly
ionized state HeII and doubly ionized state HeIII. The probabilities of
hydrogen nuclei in the configurations ${\rm H_2}$, HI and HII are thereby
specified by $x_{{\rm M}}$, $x_{{\rm HI}}$ and $x_{{\rm HII}}$, with a
normalization condition $x_{{\rm M}}+x_{{\rm HI}}+x_{{\rm HII}}=1$. The
probabilities of helium nuclei in the configurations HeI, HeII and HeIII are
specified by $x_{{\rm HeI}}$, $x_{{\rm HeII}}$ and $x_{{\rm HeII}}$ with $x_{%
{\rm HeI}}+x_{{\rm HeII}}+x_{{\rm HeIII}}=1$.

The energy of an atomic configuration is equal to the energy of bound
electrons in it, and is determined by the ionization energy and dissociation
energy. For hydrogen and helium, their configuration energies are defined by 
\begin{eqnarray}
E_{\rm H_2} & = & -2\chi_{\rm HI} - \chi_{\rm H_2}
\nonumber\\
E_{\rm HI} & = & -\chi_{\rm HI}
\nonumber\\
E_{\rm HII} & = & 0
\nonumber\\
E_{\rm HeI} & = & -\chi_{\rm HeI}-\chi_{\rm HeII}
\nonumber\\
E_{\rm HeII} & = & -\chi_{\rm HeII}
\nonumber\\
E_{\rm HeIII} & = & 0
\end{eqnarray}
The energies of ionization and dissociation $\chi _{{\rm H_2}}$, $\chi _{%
{\rm HI}}$, $\chi _{{\rm HeI}}$, $\chi _{{\rm HeII}}$ for H and He have
values 4.48, 13.598, 24.587, and 54.416 in eV. The energy of a fully ionized
configuration is equal to zero because it has no bound electron. The
configurational energy of ${\rm H_2}$ specified by $E_{{\rm H_2}}$ is equal
to the energy of two electrons bound in ${\rm H_2}$.

The probabilities of atomic nuclei in various configuration are alternative
quantities for describing ionization and dissociation degrees. The number of
atomic nuclei in each configuration is equal to the number of atomic nuclei
multiplied by the corresponding probability. The number of free electrons is
thus given for a H-He mixture by 
\begin{equation}
N_{{\rm e}}=N_{{\rm H}}x_{{\rm HII}}+N_{{\rm He}}(x_{{\rm HeII}}+2x_{{\rm %
HeIII}})\;. 
\end{equation}

\subsection{The equation for thermal equilibrium}

A H-He mixture in thermal equilibrium is described by its temperature $T$,
volume $V$ and the number of hydrogen nuclei $N_{{\rm H}}$ and that of
helium nuclei $N_{{\rm He}}$. Following the conservation equation for the
electron number (see Luo 1994) we have an equation for thermal equilibrium 
\begin{equation}
N_{{\rm H}}+2N_{{\rm He}}=C_{{\rm e}}T^{3/2}VF_{1/2}(\lambda -\epsilon _{%
{\rm cc}}/kT)+N_{{\rm H}}(x_{{\rm M}}+x_{{\rm HI}})+N_{{\rm He}}(2x_{{\rm HeI%
}}+x_{{\rm HeII}}) 
\end{equation}
with a constant $C_{{\rm e}}=(\sqrt{2}/\pi ^2)(km_{{\rm {e}}}/\hbar
^2)^{3/2} $ and the Fermi-Dirac integral $F_{1/2}$.

Of equation (19) the left-hand side counts the total number of electrons
from hydrogen and helium while the right hand side represents the
distribution of electrons in molecular, atomic, ionic and free states. On
the right-hand side, the first term is the number of free electrons, the
second term counts the bound electrons in molecular and atomic
configurations of hydrogen, and the third term counts those in the atomic
and ionic configurations of helium.

Equation (19) has an unique unknown variable which is the electron
degeneracy $\lambda $. The probabilities of nuclei in various configurations 
$x_{{\rm M}}$, $x_{{\rm HI}}$, $x_{{\rm HeI}}$ and $x_{{\rm HeII}}$ are all
its explicit functions according to Sects. 3.3 and 3.4. The Coulomb coupling
energy for each free electron $\epsilon _{{\rm cc}}$ is defined in Sect.
2.2. It is also an explicit function of the electron degeneracy $\lambda $
by means of the free electron number $N_{{\rm {e}}}$. As a result, the
equation for thermal equilibrium, equation (19), can be solved for the
electron degeneracy $\lambda $ as well as other equilibrium quantities $x_{%
{\rm M}}$, $x_{{\rm HI}}$, $x_{{\rm HeI}}$, $x_{{\rm HeII}}$ and $N_{{\rm e}%
} $.

\subsection{Equilibrium quantities of helium}

Helium has two degrees of ionization and thus has three configurations HeI,
HeII and HeIII. The probabilities of atomic nuclei in various configurations
can be determined by the formalism of the atomic configuration probability
(see Luo 1994). The grand partition function of electrons in an atomic
configuration $|NE>$ is defined by ${\cal Z}=gW\exp (N\lambda -E/kT)$ with
the statistical weight $g$, the number of bound electrons $N$, the
configuration energy $E$ (i.e. the energy of bound electrons) and the
occupation probability $W$.

As a result, we have the grand partition functions for various
configurations of helium 
\begin{eqnarray}
{\cal Z}_{\rm HeI} & = & W_{\rm HeI} \exp [2 \lambda-E_{\rm HeI} /kT]
\nonumber\\
{\cal Z}_{\rm HeII}&  =&  2 W_{\rm HeII} \exp [\lambda-E_{\rm HeII} /kT]
\\
{\cal Z}_{\rm HeIII}&  = & 1
\nonumber
\end{eqnarray}
which result from $g=1$ and $N=2$ for HeI and $g=2$ and $N=1$ for HeII. The
fully ionized configuration HeIII has its grand partition function equal to
unity because it has no bound electron and is characterized by $N=0$, $E=0,$ 
$g=1$ and $W_{{\rm HeIII}}=1$. The occupation probabilities $W_{{\rm HI}}$
and $W_{{\rm HeI}}$, which account for pressure ionization, are formulated
in Sect. 2.2.

The probabilities of helium nuclei in various configurations are therefore
given by 
\begin{eqnarray}
x_{\rm HeI} &=& {\cal Z}_{\rm HeI} / {\cal Z}_{\rm He}
\nonumber\\
x_{\rm HeII} &=& {\cal Z}_{\rm HeII} / {\cal Z}_{\rm He}
\nonumber\\
x_{\rm HeIII} &=& {\cal Z}_{\rm HeIII} / {\cal Z}_{\rm He}
\\
{\cal Z}_{\rm He} &=& {\cal Z}_{\rm HeI}+{\cal Z}_{\rm HeII}+{\cal Z}_{\rm 
HeIII}
\nonumber
\end{eqnarray}
where ${\cal Z}_{{\rm He}}$ is the total grand partition function of
elections for He. They satisfy a normalization condition $x_{{\rm HeI}}+x_{%
{\rm HeII}}+x_{{\rm HeIII}}=1$.

\subsection{Equilibrium quantities of hydrogen}

We include three configurations ${\rm H_2}$, HI and HII for hydrogen in the
EOS formalism. Because of the inclusion of molecular hydrogen the formalism
of the probability of hydrogen nuclei has a different form from those of
helium. We first give the probabilities of hydrogen nuclei in HI and HII for
which no molecular state is taken into account 
\begin{eqnarray}
y_{\rm HI} &=& {\cal Z}_{\rm HI} / {\cal Z}_{\rm H}
\nonumber\\
y_{\rm HII} &=& {\cal Z}_{\rm HII} / {\cal Z}_{\rm H}
\\
{\cal Z}_{\rm H} &=& {\cal Z}_{\rm HI}+{\cal Z}_{\rm HII}
\nonumber
\end{eqnarray}
with normalization $y_{{\rm {HI}}}+y_{{\rm {HII}}}=1$. The grand partition
functions of electrons in HI and HII are defined by 
\begin{eqnarray}
{\cal Z}_{\rm HI} &=& 2 W_{\rm HI} \exp[\lambda-E_{\rm HI}/kT]
\nonumber\\
{\cal Z}_{\rm HII} &=& 1
\end{eqnarray}
which are the results of $g=2$ and $N=1$ for HI, and the fully ionized
configuration for HII.

When we introduce the probability of hydrogen nuclei in the molecular
configuration $x_{{\rm {M}}}$, the exclusion of hydrogen nuclei in the
molecular configurations leads to the actual probabilities of hydrogen
nuclei in configurations HI and HII 
\begin{eqnarray}
x_{\rm HI} &=& y_{\rm HI} (1-x_{\rm M})
\nonumber\\
x_{\rm HII} &=& y_{\rm HII} (1-x_{\rm M}) .
\end{eqnarray}
For the dissociation of molecular hydrogen, the pressure equilibrium
constant defined by Vardya (1960) leads to 
\begin{equation}
K(T)={\frac{P_{{\rm HI}}^2}{P_{{\rm H_2}}}}={\frac{(kTN_{{\rm HI}}/V)^2}{%
kTN_{{\rm H_2}}/V}}={\frac{x_{{\rm HI}}^2}{x_{{\rm M}}}}{\ \frac{2kTN_{{\rm H%
}}}V} 
\end{equation}
with the probabilities of hydrogen nuclei $x_{{\rm M}}$ in ${\rm H_2}$ and $%
x_{{\rm HI}}$ in HI. Equation (25) has employed the number of hydrogen
molecules $N_{{\rm H_2}}=N_{{\rm H}}x_{{\rm M}}/2$ and the number of
hydrogen atoms $N_{{\rm HI}}=N_{{\rm H}}x_{{\rm HI}}$.

With the normalization conditions $x_{{\rm M}}+x_{{\rm HI}}+x_{{\rm HII}}=1$
and $y{\rm _{{HI}}}+y_{{\rm {HII}}}=1$, Eqs. (24) and (25) lead to a
solution for the probability of hydrogen nuclei in the molecular
configuration 
\begin{equation}
x_{{\rm M}}=1+R-\sqrt{R^2+2R} 
\end{equation}
with 
\begin{equation}
R={\frac{K(T)}{4{y_{{\rm HI}}^2kT}(N_{{\rm H}}/V)}}. 
\end{equation}
The Vardya's (1960) expression for the pressure equilibrium constant can be
written as 
\begin{equation}
K(T)=P_0\exp [-(D_1/kT)+(D_2/kT)^2-(D_3/kT)^3] 
\end{equation}
with a pressure constant $P_0=3.4159\times 10^{12}{\rm dyn/cm^2}$. The
constants $D_1$, $D_2$ and $D_3$ take values 4.92516, 0.237047 and 0.148409
in eV.

Temperature ionization and pressure ionization of hydrogen atoms have fatal
effects on the dissociation of molecular hydrogen. The reduction of the
number of hydrogen atoms rapidly decreases the number of hydrogen molecules.
In the above formalism, the decrease of $y_{{\rm HI}}$ reduces $x_{{\rm M}}$
in terms of $x_{{\rm M}}\approx 1/2R$ for $R\gg 1$ according to Eqs. (26)
and (27).

\subsection{Thermal equilibrium calculation}

We finally established an algorithm for calculating thermal equilibrium
quantities from Eqs. (19)-(28). Equations (21), (24) and (26) define the
probabilities of atomic nuclei in various configurations as functions of the
degeneracy $\lambda $. The equation for thermal equilibrium, equation (19),
therefore becomes an enclosed equation with one variable $\lambda $ for
which can be solved. The iteration method is an effective and fast way to
solve equation (19) in most region of the $\rho -T$ domain except in the
region where strong pressure ionization takes place. In contrast, the
bisection method is powerful and can always arrive at unique convergent
solutions in the whole $\rho -T$ domain.

The thermal equilibrium of a mixture can be well revealed by its mean
molecular weights in a $\rho -T$ domain. The mean molecular weight $\mu $ is
defined as the mean weight per particle in atomic mass unit, i.e. $\mu =\rho
VN_{{\rm A}}/N$ with the total number of particles $N$ in the mixture. We
thus have the mean molecular weights $\mu $ for a H-He mixture in terms of 
\begin{equation}
{\frac 1\mu }={\frac X{A_{{\rm H}}}}(1-x_{{\rm M}}/2+x_{{\rm HII}})+{\frac
Y{A_{{\rm He}}}}(1+x_{{\rm HeII}}+2x_{{\rm HeIII}}). 
\end{equation}
It is the result of the degrees of ionization and dissociation through the
probabilities of atomic nuclei in various configurations.

We performed a calculation of thermal equilibrium for a H-He mixture with
abundances $X/Y=0.7/0.3$. The reciprocals of the resulted mean molecular
weights $1/\mu $ are plotted in Fig. 2 for a $\rho -T$ domain. It shows
three slopes for low densities $\log \rho <0$ from low temperatures to high
ones. They correspond to the dissociation of hydrogen molecules, the
ionization of hydrogen atoms and that of helium atoms respectively. There
are other three steep slopes in the density range $0<\log \rho <2$ for low
temperatures $\log T<5$, which are the results of pressure ionization of
hydrogen atoms and helium atoms. The surface plot of Fig. 2 shows that the
minimum and maximum values of $1/\mu $, 0.422 and 1.614, correspond
respectively to the gases of completely neutral helium and molecular
hydrogen and those of completely ionized mixture.

\begin{figure}[h]
\vbox to 2.1in{\rule{0pt}{0in}}
\includegraphics{fig_2.eps}
\caption{\small
The reciprocals of the mean molecular weight $1/\mu $ in a $\rho -T$
domain for a H-He mixture $X/Y=0.7/0.3$.
}
\end{figure}

In addition, Fig. 2 reveals that the densities for pressure ionization of
hydrogen are located between 1 to 10 ${\rm g/cm^3}$ at very low temperatures
while those of helium between 10 to 100 ${\rm g/cm^3}$. The densities for
pressure ionization are almost independent of temperatures, and pressure
ionization take place in very narrow density ranges. However, we realize
that the occupation probability by equation (16) is formulated especially
for cool-dense fluids. Although it can produce pressure ionization for
partial ionized gases at high temperatures, it underestimates the effect of
pressure ionization by the Stark effect for $\log T>4.5$.

\section{Thermodynamic functions of the EOS}

\subsection{Pressure and internal energy}

The EOS uniquely determines the pressure $P(\rho ,T,X)$ and internal energy $%
E(\rho ,T,X)$ as well as other thermodynamic quantities as functions of
density $\rho $, temperature $T$ and element abundance $X,Y$. The pressure $%
P $, the internal energy density per unit mass $E/\rho V$ and their
derivatives with respect to $\rho $ and $T$ are required to dictate the
mechanical and thermal equilibria of stars and planets. In this section, we
present the comprehensive expressions for the pressure $P$ and internal
energy $E$ as functions of $T$, $N_{{\rm H}}$ and $N_{{\rm He}}$ which are
related with $T$, $\rho $ and $X$ by equation (1).

We include the following contributions to the pressure and internal energy:
(1) the classical particle (molecules, atoms and ions) and the free
electrons as Fermi-Dirac particles; (2) Coulomb coupling for $P_{{\rm cc}}$
and $E_{{\rm cc}}$; (3) pressure ionization for $P_{{\rm pi}}$ and $E_{{\rm %
pi}}$. The internal energy also include (4) the configurational terms of
atoms, ions and molecules $E_{{\rm conf}}$ and (5) the vibrational and
rotational terms of hydrogen molecules $E_{{\rm {rv}}}$.

As a result, the pressure and internal energy for a H-He mixture is written
as 
\begin{equation}
P=kT{\frac{N_{{\rm c}}}V}+{\frac 23}kT{\frac{N_{{\rm e}}}V}{\ \frac{%
F_{3/2}(\lambda _{{\rm e}})}{F_{1/2}(\lambda _{{\rm e}})}}+P_{{\rm cc}}+P_{%
{\rm pi}} 
\end{equation}
\begin{equation}
E=\frac 32kTN_{{\rm c}}+kTN_{{\rm e}}{\frac{F_{3/2}(\lambda _{{\rm e}})}{%
F_{1/2}(\lambda _{{\rm e}})}}+E_{{\rm cc}}+E_{{\rm pi}}+E_{{\rm conf}}+E_{%
{\rm rv}} 
\end{equation}
with the number of the classical particles $N_{{\rm c}}=N_{{\rm H}}(1-x_{%
{\rm M}}/2)+N_{{\rm He}}$ and the number of free electrons $N_{{\rm e}}$
given by equation (18). In the Fermi-Dirac integrals ${F_{1/2}}$ and ${%
F_{3/2}}$, the effective term of the electron degeneracy $\lambda _{{\rm e}%
}=\lambda -\epsilon _{{\rm cc}}/kT$ is determined by the electron degeneracy 
$\lambda $ and the Coulomb coupling energy for the free electron $\epsilon _{%
{\rm cc}}$. The first and second terms in Eqs. (30) and (31) represent the
ideal contributions from the classical particles and the free electrons.

\subsection{Energy of configurations}

The configurational energy comes from the energy of bound electrons in
molecules, atoms and ions. This energy is determined by the numbers of
nuclei in various configurations and the energies of various configurations
are defined by equation (17). For a H-He mixture, the total energy from
various configurations is given by 
\begin{eqnarray}
E_{\rm conf} & =  &
N_{\rm H} (x_{\rm M} E_{\rm H_2}/2 + 
x_{\rm HI} E_{\rm HI} +x_{\rm HII} E_{\rm HII} - E_{\rm H_2}/2) 
\nonumber \\
& + & N_{\rm He} (x_{\rm HeI} E_{\rm HeI} + 
x_{\rm HeII} E_{\rm HeII} +x_{\rm HeIII} E_{\rm HeIII} - E_{\rm HeI})
\end{eqnarray}
The last terms in two brackets, i.e. ($-E_{{\rm H_2}}/2$) and ($-E_{{\rm HeI}%
}$), are added to retain a zero point of the configurational energy in the
case of 100\% molecular hydrogen and neutral helium.

\subsection{Rotational and vibrational energies of hydrogen molecules}

The rotational and vibrational states of hydrogen molecules (${\rm H_2}$)
have contribution to the internal energy. The characteristic temperature of
its rotation is $\Theta _{{\rm r}}=85.38{\rm K}$ and that of its vibration
is $\Theta _{{\rm v}}=5987{\rm K}$ (see Gopal 1974). For temperatures $T\gg
\Theta _{{\rm r}}$ which is our interest for cool dense fluids, the rotation
of ${\rm H_2}$ is sufficiently excited and its rotational energy reaches the
classical value of $kT$. In contrast, the vibration of hydrogen molecules
cannot be excited to classical states at temperatures $T<\Theta _{{\rm v}}$
or $T\sim \Theta _{{\rm v}}$. Therefore, we use the Einstein function (see
Gopal 1974) for the vibrational energy of ${\rm H_2}$.

As a result, the contribution from the rotation and vibration of ${\rm H_2}$
to the internal energy is given by 
\begin{equation}
E_{{\rm rv}}=kTN_{{\rm H_2}}+\left[ {\frac{k\Theta _{{\rm v}}}{e^{\Theta _{%
{\rm v}}/T}-1}}+{\frac 12}k\Theta _{{\rm v}}\right] N_{{\rm H_2}} 
\end{equation}
with the number of hydrogen molecules $N_{{\rm H_2}}=N_{{\rm H}}x_{{\rm M}%
}/2 $. In equation (33) the first term results from the rotation of ${\rm H_2%
}$ while the second term from its vibration. The factor $E_{{\rm rv}}/kTN_{%
{\rm H_2}}$ is sensitive to temperatures around $T\sim \Theta _{{\rm v}}$.

\subsection{Contribution from Coulomb coupling}

The contribution from Coulomb coupling to the pressure and internal energy
is calculated from the Coulomb coupling energy contributed to the free
electrons. Equation (7) defines the average energy from Coulomb coupling for
each free electron $\epsilon _{{\rm cc}}$. According to the formalism of the
EOS formulated by Luo (1994), the pressure from Coulomb coupling is given by 
\begin{equation}
P_{{\rm cc}}=-\frac{N_{{\rm e}}}V\left( {\frac{\partial \epsilon _{{\rm cc}} 
}{\partial \ln V}}\right) _{TN}={\frac{N_{{\rm e}}\epsilon _{{\rm cc}}}V}{\
\frac 13}\left( {\frac{\partial \ln g}{\partial \ln \Gamma }}\right) 
\end{equation}
and the internal energy by 
\begin{equation}
E_{{\rm cc}}=N_{{\rm e}}\epsilon _{{\rm cc}}-N_{{\rm e}}\left( {\frac{%
\partial \epsilon _{{\rm cc}}}{\partial \ln T}}\right) _{VN}=N_{{\rm e}%
}\epsilon _{{\rm cc}}\left( {\frac{\partial \ln g}{\partial \ln \Gamma }}%
\right) 
\end{equation}
which are the results of equation (7). The derivative factor in Eqs. (34)
and (35) has been given by equation (10) and is a function of the Coulomb
coupling parameter $\Gamma $ and the average nuclear charge $Z$. A relation $%
P_{{\rm cc}}=E_{{\rm cc}}/3V$ is retained for the Coulomb coupling from Eqs.
(34) and (35).

\subsection{Contribution from pressure ionization}

Pressure ionization is caused by interparticle interactions and has its
contributions to the pressure $P$ and internal energy $E$. Its contributions
have been formulated as the results of the occupation probability $W$ of
bound atomic configurations (see Luo 1994) in terms of 
\begin{equation}
P_{{\rm {pi}}}=\frac 12\sum_jkT\frac{N_j}V\left( \frac{\partial \ln W_j}{%
\partial \ln V}\right) _{TN_j},
\end{equation}
\begin{equation}
E_{{\rm {pi}}}=\frac 12\sum_jkTN_j\left( \frac{\partial \ln W_j}{\partial
\ln T}\right) _{VN_j},
\end{equation}
where the sums go over all bound configurations for $j=\{{\rm H_2,HI,HeI,HeII%
}\}$ with the species number $N_j$. The factor (1/2) is introduced to take
account of the fact that the interparticle interactions responsible for
pressure ionization have actually been counted twice for each atom in the
summation.

The fully ionized ions, e.g. HII and HeIII, have no contribution because
they have no bound systems interacting with surroundings. The contribution
from molecular hydrogen should be included in that from hydrogen atoms since
its pressure dissociation results from pressure ionization of atomic
hydrogen according to Eqs. (26) and (27). As a result, the contribution from
pressure ionization to the pressure for a H-He mixture is given by 
\begin{eqnarray}
P_{\rm pi} & =  &
{1 \over 2} kT {N_{\rm H} \over V} (x_{\rm M}+x_{\rm HI})
{\left( \partial \ln W_{\rm HI} \over \partial \ln V \right)}
\nonumber \\
& + &
{1 \over 2} kT {N_{\rm He} \over V} 
\left[
x_{\rm HeI} \left( \partial \ln W_{\rm HeI} \over \partial \ln V \right) +
x_{\rm HeII} \left( \partial \ln W_{\rm HeII} \over \partial \ln V \right) 
\right]
\end{eqnarray}
and that to the internal energy given by 
\begin{eqnarray}
E_{\rm pi} & =  &
{1 \over 2} kT N_{\rm H}  (x_{\rm M}+x_{\rm HI})
{\left( \partial \ln W_{\rm HI} \over \partial \ln T \right)}
\nonumber \\
& + &
{1 \over 2} kT N_{\rm He} 
\left[
x_{\rm HeI} \left( \partial \ln W_{\rm HeI} \over \partial \ln T \right) +
x_{\rm HeII} \left( \partial \ln W_{\rm HeII} \over \partial \ln T \right) 
\right] .
\end{eqnarray}

The derivative terms in Eqs. (38) and (39) are determined by the formalism
of the occupation probability $W_j$. They have analytical expressions for $%
j=\{{\rm HI,HeI,HeII}\}$ 
\begin{equation}
\left( {\frac{\partial \ln W_j}{\partial \ln V}}\right) _{TN_j}={\frac{4\pi }%
3}r_j^3\left[ {\frac{N_{{\rm H}}+N_{{\rm H}}}V}+{\frac 23}C_{{\rm e}%
}T^{3/2}F_{3/2}(\lambda _1)h(\lambda _1)\right] 
\end{equation}
\begin{equation}
\left( \frac{\partial \ln W_j}{\partial \ln T}\right) _{VN_j}={\frac{4\pi }3}%
r_j^3C_{{\rm e}}T^{3/2}F_{3/2}(\lambda _1)\left[ h(\lambda _1)-1\right] 
\end{equation}
as the results of equation (16). The degeneracy parameter of the acting
electrons $\lambda _1$ is defined by equation (15). The factor $h(\lambda
_1) $ is a function of $\lambda _1$ and has a form 
\begin{equation}
h(\lambda _1)={\frac{3F_{1/2}^2(\lambda _1)}{F_{3/2}(\lambda
_1)F_{-1/2}(\lambda _1)}} 
\end{equation}
which has asymptotic values $h=1$ for $\lambda _1\ll 0$ and $h=5/3$ for $%
\lambda _1\gg 0$.

\subsection{Discussion}

In this section, we archive a complete algorithm for calculating the
pressure and internal energy with Eqs. (30)-(42). All of the expressions
have explicit and simple form although some approximations have been
adopted. The first approximation is the use of one-component plasma model
for Coulomb coupling in Sect. 2.1. However, it is a good approximation when
the corrections from partial ionization and element mixture are added. And
it is valid for a very wide range of the Coulomb coupling parameter with $%
0\le \Gamma \le 160$. The second one is the model of pressure ionization
established in Sect. 2.2. It is based on the concept of the pressure impact
of acting particles on the acted-upon atom, and represents the ionization of
atoms by the high pressure of matter. Its simplicity allows to retain the
analytical expressions for the contribution from pressure ionization to the
pressure and internal energy.

The algorithm of the EOS established in Sect.3 and 4 is based on the
chemical picture for mixed elements. It allows to easily include more
elements for their contribution to the EOS. The formalism for hydrogen can
be developed for the elements with molecules while that for helium can be
developed for the elements with multielectron atoms.

\section{Results}

\subsection{Contribution from nonideal effects}

Pressure ionization and Coulomb coupling lead to the nonideal effects on the
EOS. They not only increase the ionization degrees, but also contribute to
thermodynamic functions. In Sect. 4, the formalism for the EOS separates
their contribution terms from those ideal terms of classical particles and
free electrons. This allows to examine the importance of pressure ionization
and Coulomb coupling for their contribution to the pressure and internal
energy.

In spite of the reliability of our EOS for very cool-dense fluids, we
performed the calculation upto very high densities for low temperatures $%
(T\ge 10^2K)$. That reaches the states of molecular solid and metallic solid
(see Kerley 1972). The simplicity of the EOS formalism allows to easily
obtain the solution and to roughly investigate the properties of matter from
hot gases, to cool-dense fluids, and to solids.

We first examine the contributions from pressure ionization and Coulomb
coupling to the pressure and internal energy for a H-He mixture with $%
X/Y=0.7/0.3$. Their fractions over total pressure and internal energy are
plotted in Fig. 3 and Fig. 4 for a $\rho -T$ domain. They are actually
formulated by $(P_{{\rm pi}}+P_{{\rm cc}})/P$ and $(E_{{\rm pi}}+E_{{\rm cc}%
})/E$ . Equations (34), (36), (38) and (39) indicate that pressure
ionization has positive contributions to the pressure and internal energy
while Coulomb coupling has negative contributions to them.

\begin{figure}[h]
\vbox to 2.1in{\rule{0pt}{0in}}
\includegraphics{fig_3.eps}
\caption{\small
The fractions of the contribution from pressure ionization and
Coulomb coupling over the total pressure in a $\rho -T$ domain for a mixture 
}
\end{figure}

\begin{figure}[h]
\vbox to 2.1in{\rule{0pt}{0in}}
\includegraphics{fig_4.eps}
\caption{\small
Same as Fig. 3, but over the total internal energy.
}
\end{figure}

Figures 3 and 4 reveal that the contribution from pressure ionization
increases the pressure $P$ and internal $E$ before pressure ionization takes
place. In contrast, the contribution from Coulomb coupling reduces $P$ and $%
E $ when pressure ionization occurs and a great fraction of charged
particles are produced. In addition, the contribution from pressure
ionization dominates both pressure and internal energy in a density range of 
$-1\le \log \rho \le 1$ for temperatures $\log T\le 4.5$. The contribution
from Coulomb coupling can reduce the pressure and internal energy by tens
percent just when pressure ionization takes place at very low temperatures.

In Figs. 5 and 6 the pressure $P$ and internal energy density $E/V\rho $ are
plotted on isotherms in a density range $-2\le \log \rho \le 2$ in solid
lines. Together plotted in dotted lines are the ideal terms, i.e., $P-P_{%
{\rm pi}}-P_{{\rm cc}}$ and $(E-E_{{\rm pi}}-E_{{\rm cc}})/V\rho $ which are
the pressure and internal energy excluding the contributions from pressure
ionization and Coulomb coupling. From Fig. 5 and Fig. 6 we see the following
points. (1) The densities of pressure ionization are located in a very
narrow range. There the ideal term of free electrons increases the pressure
and internal energy rapidly and significantly. This is the result of the
fast production of free electrons due to pressure ionization. (2) The
contribution from pressure ionization starts to play a role and increases
the pressure and internal energy far before pressure ionization takes place.
This results from the interaction among atoms when they are close to each
other. (3) The contribution from Coulomb coupling plays a role just after
pressure ionization takes place. A great fraction of charged particles are
produced and the resulted Coulomb coupling from them reduces both pressure
and internal energy. (4) The increase of the pressure and internal energy by
the contribution from pressure ionization before occurring of pressure
ionization and the decrease by that from Coulomb coupling after occurrence
of pressure ionization smooth the variation of both pressure and internal
energy in the region of pressure ionization, in spite of the fact that the
fast production of free electrons by pressure ionization increase the ideal
pressure and internal energy steeply and significantly.

\begin{figure}[h]
\vbox to 2.1in{\rule{0pt}{0in}}
\includegraphics{fig_5.eps}
\caption{\small
Pressure isotherms (in unit dyn/cm$^2$) in solid lines, compared with those of the ideal
terms in dotted lines. The isotherms are (from bottom to top): $\log
T=2.0, 2.5, 3.0, 3.5, 4.0, 4.5, 5.0,  5.5, 6.0$.
}
\end{figure}
\begin{figure}[h]
\vbox to 2.1in{\rule{0pt}{0in}}
\includegraphics{fig_6.eps}
\caption{\small
Same as Fig 5 but for internal energy density isotherms (in unit erg/g).
}
\end{figure}

As a result, we may conclude that the contributions from pressure ionization
and Coulomb coupling smooth the variation of the pressure and internal
energy in the region of pressure ionization even when pressure ionization
takes place in a very narrow density range.

\subsection{Comparison of the EOS}

The EOS formulated in this paper has simple and explicit expressions for its
practical use. However, its reliability requires comparison with the EOS
data calculated by other EOS. Of many EOSs commonly used in astrophysics we
choice the EOS calculated by Saumon et al. (1995) to compare our EOS. Their
EOS is based on careful study of nonideal interaction and new physical
treatment of partial dissociation and ionization by both pressure and
temperature effects (see Saumon and Chabrier 1991,1992). They also made
comparison with other EOS for a wide $\rho -T$ domain, and gave detailed
discussion and analysis.

In Fig. 7 and Fig. 8 we plot the pressure and internal energy density on
isotherms from our EOS and the Saumons' EOS for a pure hydrogen fluid. The
difference at low density and high temperature comes from that our EOS
includes the contribution from radiation while the Saumon's does not. To fit
with Saumons' EOS on the internal energy density at very low temperatures,
we have to ignore the zero temperature term of vibrational energy of
molecular hydrogen, i.e., the term $(1/2)k\Theta _{{\rm v}}$ in equation
(33). Comparison in the pressure in Fig. 7 shows difference mainly at low
temperatures $\log T\le 4.5$ in the density range $-1\le \log \rho \le 1$.
Comparison in the internal energy density in Fig. 8 also shows the
difference in the same region. These differences come from the different
formalisms for the models of pressure ionization and Coulomb coupling. In
spite of this, both of them prove the same tendency of the pressure and
internal energy from neutral fluids to ionized plasmas.

\begin{figure}[h]
\vbox to 2.1in{\rule{0pt}{0in}}
\includegraphics{fig_7.eps}
\caption{\small
Comparison of pressure isotherms (in unit dyn/cm$^2$) between our EOS (solid lines) and
the Saumons' EOS (circles) for a pure hydrogen fluid. The isotherms are
(from bottom to top): $\log T$ $=2.5, 2.9, 3.3, 3.7, 4.1, 4.5, 4.9, 5.3, 5.7$.
}
\end{figure}
\begin{figure}[h]
\vbox to 2.1in{\rule{0pt}{0in}}
\includegraphics{fig_8.eps}
\caption{\small
Same as Fig. 7 but for internal energy density isotherms (in unit erg/g).
}
\end{figure}
Another distinguished difference is revealed in the internal energy isotherm
at temperature $\log T=4.5$ in Fig. 8. The difference covers a density range 
$-4\le \log \rho \le -1$ and is specified by $\Delta \log E=0.07$ or $17$
percent. In this region our EOS results are more close to the EOS of MHD of
which a difference is found with about the same amount (see Saumon et al.
1995). The difference is argued by Saumon et al. (1995) to come from a
higher degree of ionization caused by the increase of the finite size of
atoms with temperature and density dependence.

For higher temperatures $\log T\ge 5$, our EOS has almost the same result as
Saumons' EOS in the internal energy. However, both EOSs have smaller values
than the MHD EOS. This fact has been revealed by Saumon et al. 1995. This is
because both EOSs do not take into account the Stark ionization while the
MHD EOS does and yields higher degrees of ionization for a partially ionized
gas.

\section{Conclusions}

We present a single and complete algorithm of the EOS from cool-dense fluids
($T>10^2$K) to hot gases for a H-He mixture. The physical processes of
pressure ionization and Coulomb coupling are simply modeled for its validity
from cool-dense fluids to hot gases and for mixed elements. This allows to
archive an EOS with explicit expressions which can be easily incorporated
into the modeling of stars or planets. In addition, the simple and
comprehensive algorithm also allows to incorporate more physics processes or
add more elements into the EOS formalism.

We show that the thermodynamic functions remain smooth even at the abrupt
occurrence of pressure ionization at very low temperatures. The
contributions from the Coulomb coupling and pressure ionization to the
thermodynamic function compensate the effect of the rapid increase of free
electrons when pressure ionization takes place. Although the quantum
exchange effect of electrons is not included, the resulted algorithm
produces comprehensive data of the EOS from hot gases to cool-dense fluids.

\acknowledgments 
This work was supported by the Science Foundation of Yunnan and the National
Nature Science Foundation of China. Thanks are also due to D. Saumon and H.
Chabrier for their EOS data which we accessed by internet ftp.

\end{document}